\documentclass[fleqn,10pt]{SelfArx} 

\usepackage{amsmath,amssymb,graphicx}
\usepackage{snapshot}

\definecolor{color1}{RGB}{0,0,0} 
\definecolor{color2}{RGB}{255,255,255} 

\JournalInfo{submitted: June 8, 2014} 
\Archive{ } 

\PaperTitle{Single-shot 3D sensing with improved data density} 

\Authors{Florian Willomitzer\textsuperscript{1}*, Svenja Ettl\textsuperscript{1}, Christian Faber\textsuperscript{2}, and Gerd H\"ausler\textsuperscript{1}} 
\affiliation{\textsuperscript{1}\textit{Institute of Optics, Information and Photonics, University Erlangen-Nuremberg, Staudtstr.
7/B2, 91058 Erlangen, Germany}} 
\affiliation{\textsuperscript{2}\textit{Sensor Technology and Image Processing, University of Applied Sciences Landshut, Am Lurzenhof 1, 84036 Landshut, Germany}} 
\affiliation{*\textbf{Corresponding author}: florian.willomitzer@physik.uni-erlangen.de} 

\newcommand{\keywordname}{Keywords} 

\Abstract{We introduce a novel concept for motion robust optical 3D-sensing. The concept is based on multi-line triangulation. The aim is to evaluate a large number of projected lines (high data density) in a large measurement volume with high precision. Implementing all those three attributes at the same time allows for the "perfect" real-time 3D movie camera (our long term goal). The key problem towards this goal is ambiguous line indexing: we will demonstrate that the necessary information for unique line indexing can be acquired by two synchronized cameras and a back projection scheme. The introduced concept preserves high lateral resolution, since the lines are as narrow as the sampling theorem allows, no spatial bandwidth is consumed by encoding of the lines. In principle, the distance uncertainty is only limited by shot noise and coherent noise. The concept can be also advantageously implemented with a hand-guided sensor and real-time registration, for a complete and dense 3D-acquisition of complicated scenes.}

\begin{document}

\maketitle 

\flushbottom 


\thispagestyle{empty} 

\section{Scope of work}

We will introduce a novel concept for motion robust optical 3D-sensing. The concept is based on multi-line triangulation, due to its single-shot ability. We will demonstrate how to increase the number of acquired 3D-points, by significantly increasing the number of projected lines. The upcoming indexing problem is solved by introducing one or more additional cameras. Compared to other approaches that exploit spatial line encoding, our concept does not consume space-bandwidth. So, our sensor displays the best possible lateral resolution as well as small distance uncertainty.

Optical 3D sensors are standard tools for many applications in industry, medicine, art conservation or virtual reality. The underlying reason is that 3D-data offers serious benefits, compared to 2D-images: the shape is invariant against translation or rotation of the object as well against varying illumination, texture or soiling. No wonder that there are hundreds of 3D-sensors available, all aiming for "high accuracy", "high speed", "comfortable use" or similar attributes \cite{LeachBuch11}. Surprisingly, there are only a few concepts that address the attribute "single-shot" or "motion-robust". By "single-shot", we understand that no sequentially acquired exposures are necessary for the 3D data generation. All information is supposed to be acquired within the acquisition time of one single camera frame. Such a feature would be of great importance, for two reasons: first, the object (or the sensor) is not required to stand still, which will allow for the measurement of continuously moving objects in industry or for medical applications. One even more important aspect is that the acquisition of a complete object commonly requires 3D-views from many directions. In order to avoid elaborate repositioning of the sensor, a single shot concept allows for an easy (even hand-guided) full $360 ^\circ$  3D-scan because the sensor is allowed to move continuously. Meanwhile, a few single shot concepts are known and a number of sensors based on this principle are available. However, single-shot sensors have a major drawback: without the exploitation of an additional modality, single-shot sensors are principally not able to acquire a dense 3D-point cloud.

In \cite{HaeuslerDGaO12} the question has been posed, if a single-shot sensor is theoretically capable to acquire a cloud of uncorrelated (not interpolated) 3D points with the full resolution of the camera chip. The answer is somewhat depressing: physics and information theory do not allow for such a 3D-sensor, at least not with the common configuration of triangulation, using projected patterns and a black-and-white camera, or by stereo photogrammetry. The single-shot feature has to be purchased by a reduced density of uncorrelated 3D-data.

\begin{figure}[htbp]
\centerline{\includegraphics[width=\columnwidth]{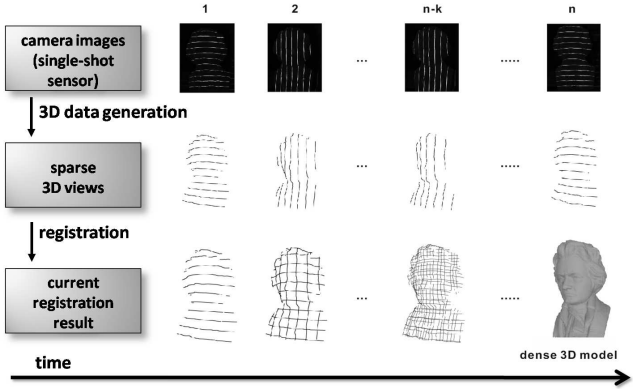}}
\caption{Basic scheme of Flying Triangulation. Data acquisition process: two perpendicularly aligned line patterns are subsequently projected onto the object surface. For each projection step, an image is taken with a camera, using a fixed triangulation angle $ \theta$. Sparse 3D views can be calculated directly from each single camera exposure. These sparse views are aligned to each other \cite{Arold14}. The data acquisition- and registration-process is done simultaneously, which allows for a visual feedback of the current registration outcome in real-time. The measurement is finished, when the resulting data display a complete and dense 3D-model. For common objects, an all around 3D-image takes about five to ten seconds. Examples are available on \cite{Osmin3D}.}
\label{MeasPrinc_FlyTri.eps}
\end{figure}

In \cite{Ettl12} we introduced a workaround, called "Flying Triangulation": a hand-guided single-shot sensor acquires a series of 3D views, each view containing only sparse data. Via a sophisticated registration algorithm, the single 3D views are aligned to each other in order to create a dense 3D-model with high lateral resolution (see Fig.~\ref{MeasPrinc_FlyTri.eps}). Eventually, the sensor needs a sequence of images for a dense model, BUT each 3D point is generated from only one exposure and is completely uncorrelated to others (within the limits of the spatial bandwidth). This allows for a continuous relative movement of object and sensor without relying on external tracking.

In the long term, we aim for the information-theoretically optimal real-time "3D-movie-camera". By this, we understand a 3D sensor which is able to acquire all kinds of object shapes in real-time with a high data density in a large depth of field. Since high object frequencies should be measurable within one single camera exposure, neither temporal nor spatial context information can be exploited. Principally, Flying Triangulation is suited for this task because it already fulfills some important criteria mentioned above. However, it lacks immediate data density and it lacks a large measurement depth (the interdependence between these quantities will be clarified in the third section).

Therefore, as starting point, the following question is of special interest: How much uncorrelated 3D points $N_{3D}$ can we acquire from one picture with $N_{pix}$ pixels? A typical Flying Triangulation sensor with a 1 Megapixel camera projects roughly 10 lines, resulting in a 3D data-efficiency of

\begin{equation}
 \eta_{3D} =  \frac{N_{3D}}{N_{pix}} \approx 1\%.
\end{equation}

The obvious question is: why do we project only 10 lines? Where is the limit to exploit the full resolution of our, say,  1 Megapixel camera? Two serious limitations make up the answer: The first is discussed in \cite{HaeuslerDGaO12}: For an artifact-free localization of the recorded line signals, the sampling theorem has to be satisfied. If, in addition,  a perspective contraction at tilted surfaces is considered, the author deduces a conservative limit of the pixel-efficiency $\eta_{3D}$ not better than 16\%. For a 1 Megapixel camera, this corresponds to 160 lines. However, current triangulation sensors, using only monochrome straight narrow lines for projection, are still far away from this limit. This is due to the second severe limitation: the problem of ambiguous line indexing. For multi-line triangulation, the projected lines have to be correctly indexed in the camera image in order to obtain the correct correspondence between signal and observation. Due to the indistinguishability of straight lines, this becomes a difficult task for bigger line numbers. 

Many approaches (a few examples will be discussed in the next section) were developed to overcome the correspondence problem. Most of them exploit temporal or lateral context information. This means that the applied signals are modified in space or time in order to become distinguishable.  As a consequence, bandwidth is consumed, that is not available anymore for high temporal or lateral resolution.

In this paper, we introduce a new method for indexing in multi-line triangulation. It does not rely on the exploitation of spatial or temporal bandwidth. The basic idea is to acquire the necessary information from \textit{several synchronized cameras}.

\section{State of the art and its limitations }

The task to resolve ambiguities of signal correspondences appears frequently in fast 3D-metrology aiming for high density of 3D data. Existing approaches can be categorized into three groups \cite{PagesSalvi03,Geng10}: spatial neighborhood, time-multiplexing and usage of an additional modality such as color. Clearly, indexing strategies exploiting the spatial neighborhood cannot produce pixel-dense uncorrelated data, whereas time-multiplexed methods are not single-shot. By "uncorrelated", we understand that each 3D point is solely evaluated with the information of one single camera pixel and contains no information of neighbored pixels. In the following, we will discuss common 3D measurement techniques. Special attention is given to the related correspondence approaches, their benefits and drawbacks.

The paradigm of a 3D-sensor with high density of 3D-data is the so called "fringe projection" \cite{Halioua84}. This principle requires at least three subsequent exposures. These are necessary, since each camera pixel will acquire information about the ambient light, the object reflectivity and the local phase (which encodes the distance). The three exposures yield data to calculate these three unknowns. With a single gray scale camera, there is no way to acquire dense 3D-data within one single shot. An exception occurs, if the bandwidth of the fringe image is at least 3 times smaller than the camera bandwidth. Then, one can decipher the phase of the fringes by a Fourier technique \cite{Takeda83}.

There are also workarounds towards a short acquisition time: sensors which need a series of exposures, but which are extremely fast, so for the user they appear to be motion robust, in practice. Those approaches are suggested, for example by \cite{Schaffer10} or \cite{Munkelt08}.

By introducing another modality such as color, a temporal sequence of exposures can be avoided. In 1993, H\"ausler and Ritter introduced the color-encoded triangulation \cite{Haeusler93}. Instead of Fringes, a continuous spectrum of wavelengths is projected. Then, the hue within each pixel can be uniquely determined, no ambiguity occurs. The sensor provides pixel-dense single shot acquisition. However, this approach displays some drawbacks: it is technically elaborous and expensive to project a bright spectrum, and the intrinsic object color may disturb the result. Furthermore, high quality color cameras are required for the decoding. Nevertheless, the method might become important, since bright sources for a wide spectrum are available nowadays \cite{Russell02}.

In the following we will not further discuss time-multi-plexing and color-coding methods, since we are interested in the more fundamental question, to what extent a "real" single shot sensor without color encoding might be possible. In this regime, most sensor principles use the spatial neighborhood to avoid ambiguity. The authors of \cite{PagesSalvi03} and \cite{Geng10} roughly categorized several methods. There are as well hybrid approaches, which combine spatial neighborhood with time multiplexing or color (see \cite{Schmalz12,Zhang02,Kawasaki08,Koninckx03} and many others) which are not further discussed here.

The most common solution is spatial encoding of the projected signal. Here, the basic mechanism is to make neighbored signals distinguishable with a unique sequence ("codeword") of signals ("letters"). In the well established De Bruin codification \cite{MacWilliams76}, different intensities (or colors) of projected stripes are used to create a cyclic pseudorandom sequence, that encrypts the side by side alignment of the projected stripe pattern. Although the correspondence problem is solved via the information of this whole sequence, the local depth is solely measured via the position of the edge between two stripes. However, as space is required to read the codeword, it is not possible to measure at close distance to edges and shaded areas. This is even more affected by long codewords.

In some approaches, a codification \textit{along} each line \cite{Artec,WXie} (rather than perpendicular to it) is used, e.g. by changing the width along the line direction. Although this approach allows to consider each line independently from the neighbored lines, it needs sufficient space in the line direction for identification. As of the space required by the locally bigger line width, the possible number of projected lines is reduced compared to non encoded lines.

Codification can also be applied in two dimensions \cite{Albitar07}. These methods can be advantageous for discontinuous surfaces, because the euclidean distance between all relevant letters is smaller.

All single-shot methods using spatial neighborhood approaches reach a comparatively high data density but have one weak point in common: They exploit lateral context information to encrypt the codeword. Sharp edges or fine 3D details cannot be measured properly. If different gray-values are used in order to shorten the codeword, the measurement of objects with texture, varying reflectivity or ambient illumination might be difficult.

Now we pose the crucial question of our paper: Can we solve the correspondence problem simply by introducing one or more \textit{additional cameras}? A passive approach is stereo photogrammetry. The basic weakness is that the object surface must display features that represent corresponding points in the two or more camera images. For applications such as virtual reality, the method may work sufficiently well \cite{ScannerKiller}, for 3D-metrology, passive stereo is not well suited: first, there is no chance to measure unstructured surfaces, second, the identification of corresponding features again needs lateral context information. For unstructured objects, there are active stereo photogrammetry approaches. The standard solution is based on the projection of a random "speckle-like" pattern. This concept is quite simple and easy to realize. However, the encoding is inefficient and also sucks up bandwidth. Hence, it does not deliver an uncorrelated point cloud.  

We avoid the drawbacks described above and present a completely different approach for the line indexing in multi-line triangulation systems. We do not want to exploit any neighborhood information: Each pixel at the camera target should have sufficient information to uniquely identify the line number and to evaluate an accurate distance.
The starting point of our considerations is Flying Triangulation. However, the resulting technique is valid for all multi-line triangulation systems in general.

\section{Ambiguities in line indexing }

\begin{figure}[t]
\centerline{\includegraphics[width=.74\columnwidth]{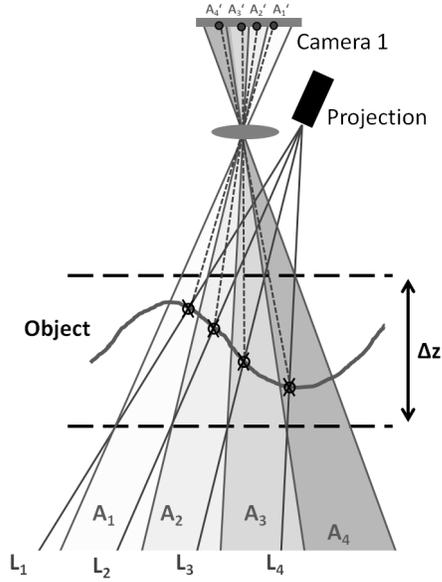}}
\caption{Present indexing approach (illustrated for 4 lines):  Inside the measurement depth $\Delta z $ each line $L_1 ,..., L_4$ is only located in its individual region of uniqueness $A_1 ,..., A_4$. Points measured inside $\Delta z $ by intersection of $L_1 ,..., L_4$ with the object surface are automatically imaged within their corresponding area $A_1' ,..., A_4'$ at the camera chip. This method yields unique indexing inside the predefined measurement depth $\Delta z $.}
\label{Indexing_FlyTri.eps}
\end{figure}

Light sectioning is an established 3D measurement principle. As mentioned, it lacks efficiency. By projecting multiple lines, the efficiency can be increased. Then, light sectioning relies on indexing strategies to decipher the correct correspondences. Due to the problem of shadowing or varying reflectivity, some lines may not be visible in the camera image. This makes indexing a complicated task which is the reason that the theoretically discussed efficiency (see section 1) is not reached yet. Flying Triangulation follows an indexing approach that is completely independent of the object surface. Even "complicated" objects with steps like human teeth can be measured. The basic idea is to dedicate for each line a unique region on the camera chip, that directly corresponds to its index (see Fig. ~\ref{Indexing_FlyTri.eps}). Inside the predefined measurement range $\Delta z $, each projected line $L_1 ,..., L_N$ will be located only in an assigned "region of uniqueness" $A_1 ,..., A_N$. Hence, surface points, measured inside $\Delta z $ by intersection with $L_1 ,..., L_N$, are automatically imaged onto the corresponding areas $A_1' ,..., A_N'$ on the camera chip. Here, the line index $n$ is determined. If this index $n$ is known, the calibration-transformation $K_C$, that maps each 3D-point $(x,y,z)$ in space to pixel coordinates $(i,j)_C$ onto the camera chip of camera $C$

\begin{equation}
 \left(\begin{array}{c}i\\j\end{array}\right)_C = K_C \left(\begin{array}{c}x\\y\\z\end{array}\right)
\end{equation}

can uniquely be inverted (see Eq.~(\ref{eq:Abbildung})).

Therefore, with the pixel coordinates $(i_S,j_S)_{C_1}$ of the signal and the determined index $n_{C_1}$, 3D data $(x_e,y_e,z_e)$ of the measured object can be evaluated via:

\begin{equation}
 \left(\begin{array}{c}x_e\\y_e\\z_e\end{array}\right) = {K_{C_1}}^{-1} \left(\begin{array}{c}i_S\\j_S\\n\end{array}\right)_{C_1}
\label{eq:Abbildung}
\end{equation}

\begin{figure}[t]
\centerline{\includegraphics[width=.8\columnwidth]{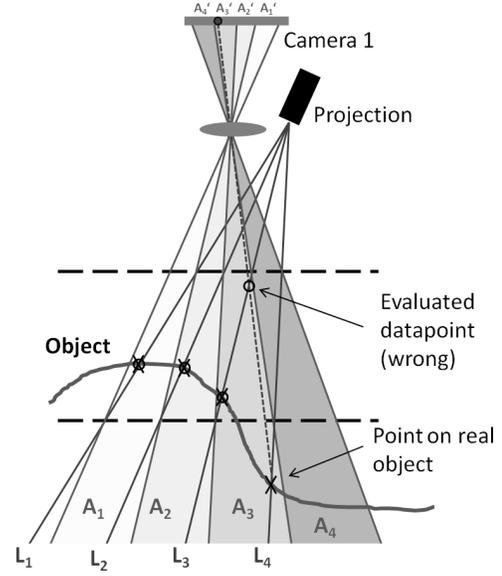}}
\caption{False line indexing (illustration). The projected line $L_4$ intersects the object surface outside the defined measurement depth. The signal from this point is located in region $A_3$ and imaged onto area $A_3'$ on the camera chip. Hence, a wrong line index will be assigned (3 instead of 4), which yields a false evaluated 3D data point.}
\label{IndexingERROR_FlyTri.eps}
\end{figure}

\begin{figure}[t!]
\centerline{\includegraphics[width=.8\columnwidth]{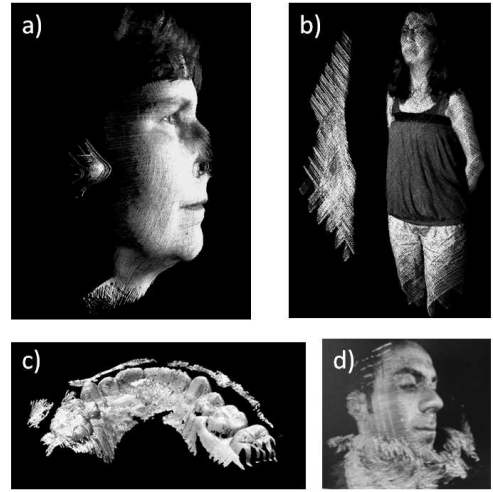}}
\caption{Examples for errors in the line indexing process: a),d): Face scanner; b): Body scanner; c) Dental scanner. The false indices yield outliers in the 3D-dataset, which may even intersect with the correctly indexed datapoints.}
\label{OutlierExamples.eps}
\end{figure}

This method yields unique indexing inside the defined measurement depth $\Delta z$. However, there is a weakness: Indexing fails if a surface point is measured outside the measurement range $\Delta z$. For hand-guided triangulation sensors, this may occur frequently. The consequence is a wrong line index which leads to false 3D data, as shown in Fig.~\ref{IndexingERROR_FlyTri.eps}: As soon as a projected line (here line $L_4$) intersects the object surface outside $\Delta z$, the signal is automatically imaged at the wrong area of uniqueness (here $A_3'$). Hence, the signal is treated like coming from a different line (here line $L_3$). This error in the line indexing results in a false 3D data point (outlier) in the final dataset. Figure~\ref{OutlierExamples.eps} shows examples for 3D models containing outliers from false indexing.

But how is this associated with our original aim for higher data density? Unfortunately, the unique measurement depth and the number of lines are closely related. For a fixed triangulation angle, both cannot be chosen independently, so there is no simple way to increase both important parameters at the same time. More projected lines will automatically result in a reduction of $\Delta z$ and outliers in the 3D dataset. 
If, as for Flying Triangulation, a registration of the 3D datasets is applied, outliers have severe consequences: 3D views with too many outliers cannot be registered, which makes an all around-view of the object impossible \cite{Arold14}.

\begin{figure}[b!]
\centerline{\includegraphics[width=.8\columnwidth]{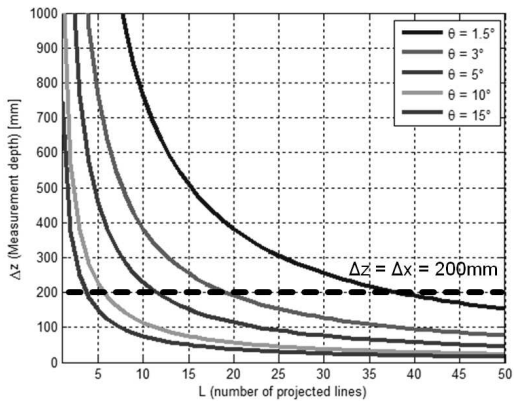}}
\caption{The relation between triangulation angle $\theta$, unique measurement depth $\Delta z$ and the number of uniquely identifiable lines for multi-line triangulation systems.  The figure illustrates the telecentric case with a lateral width of the measurement field of $\Delta x =~200~mm$. If neither lateral, nor temporal or other context information is used for the indexing process, only these solutions are possible.}
\label{Loesungsschale.eps}
\end{figure}

The mentioned relation is described by Eq.~(\ref{eq:Eindeutigkeit1}) and also Fig.~\ref{Loesungsschale.eps}. For simplification, we consider telecentric geometry for the optical setup.

\begin{equation}
L  = \frac{\Delta x}{\Delta z \cdot tan~\theta} 
\label{eq:Eindeutigkeit1}
\end{equation}

Here, $L$ is the number of projected lines, $\theta$ is the triangulation angle and $\Delta x$ is the lateral width of the measurement field (perpendicular to the line direction) which is determined by the applied lens and working distance of the sensor.

We define the "uniqueness-index" $ U $ of multi-line triangulation setups:

\begin{equation}
U = \frac{\Delta z}{\Delta x} \cdot  L \cdot { tan~\theta}
\label{eq:Eindeutigkeitslevel1}
\end{equation}

If we consider the measurement volume as cubic, means $\Delta x~=~\Delta y~=~\Delta z$, $U$ simplifies to

\begin{equation}
U = L \cdot { tan~\theta} .
\label{eq:Eindeutigkeitslevel}
\end{equation}

If no further information is exploited, only solutions of Eq.~(\ref{eq:Eindeutigkeitslevel}) and (\ref{eq:Eindeutigkeitslevel1}) for \textbf{$U\le1$} can be realized. A few examples: A typical triangulation sensor with a cubic volume of $200\times200\times200mm^3$, a 1Mpix camera and a triangulation angle of $\theta = 10^{\circ}$ displays a statistical measurement uncertainty of approximately $150\mu m$. However, due to Eq.~(\ref{eq:Eindeutigkeitslevel}) (see also Fig.~\ref{Loesungsschale.eps}), it is only possible to project 5 lines with this sensor in order to avoid outliers. If, since a higher data density is desired, $\theta$ is reduced to $\theta = 1.5^{\circ}$, approximately 40 lines could be projected but the measurement uncertainty rises to more than $1~mm$! It is obvious that data density has to be purchased by precision and measurement depth (for non cubic measurement volumes, see Eq.~(\ref{eq:Eindeutigkeitslevel1}))

Now, our goal of the first chapter can be re-defined: An increase of $U$  towards a value $\gg1$ in order to acquire a \textit{high density} of datapoints (many lines) in a \textit{large measurement depth} with \textit{high precision} (large triangulation angle). In order to accomplish this, we have to tap a new source of information!

\section{Robust line indexing via signal back-projection }

\begin{figure}[b!]
\centerline{\includegraphics[width=.8\columnwidth]{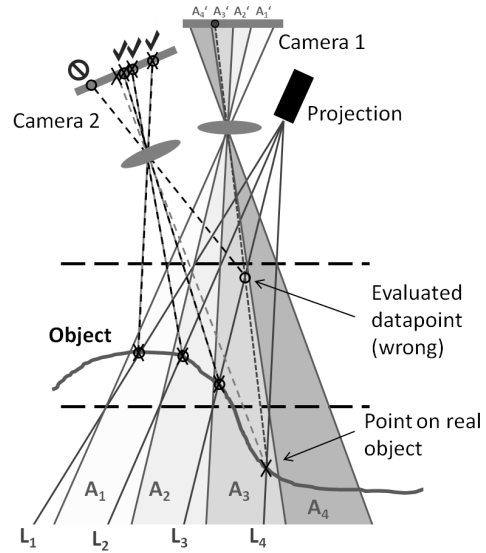}}
\caption{Is a measured surface point inside the measurement volume? The evaluated data points are back projected onto the chip of camera 2. The position of the back projection is compared to the real signal on the chip. If the signal and the back projection match, the evaluated 3D point is correctly indexed.}
\label{StereoMethod.eps}
\end{figure}

Indexing errors can be interpreted as lack of information. In this section we explain the principle of a novel indexing approach which exploits information simultaneously acquired by one or more additional cameras. In contrast to other indexing methods it employs neither color encoding nor spatial neighborhood or time-multiplexing. We will further discuss how the introduced approach can be enhanced towards a 3D movie camera. Another approach which in fact exploits context information but does not require additional hardware is also part of our current work and will be shown in \cite{Ettl14}.

It was discussed that the multi-line triangulation sensor will deliver false line indexing if a surface point is outside of the unique measurement depth $\Delta z$. So, it would already be helpful if we could decide if a certain point is inside or outside $\Delta z$: false data could be suppressed. This additional information is provided by a second camera. In many of our existing sensor setups, a second camera is already implemented for the acquisition of color texture \cite{WillomitzerDGaO11}. We emphasize that color is not used here as an additional modality to encode the depth. We also emphasize, that our approach does not rely on the conventional procedures of active stereo photogrammetry. Our solution with two cameras is as follows:

\begin{itemize}

\itemsep1em

\item Evaluation of 3D data with camera 1 (data could be falsely indexed!)

\item Back projection of the evaluated 3D data onto the chip of camera 2

\item Direct comparison of the back projection positions with the observed intensity on the chip of camera 2

\end{itemize}

The principle is illustrated in Fig.~\ref{StereoMethod.eps}. 3D points, generated from signals inside the unique measurement depth $\Delta z$ correspond to real points on the object surface. Since the transformation $K_{C_2}$ between the 3D space and the pixel-coordinates $(i,j)_{C_2}$ of camera 2 is known, each evaluated 3D data point $(x_e,y_e,z_e)$ can be back projected onto the chip of camera 2. We obtain the pixel-coordinates $(i_B,j_B)_{C_2}$ for the back projections on the chip of camera 2:

\begin{equation}
 \left(\begin{array}{c}i_B\\j_B\end{array}\right)_{C_2} = {K_{C_2}} \left(\begin{array}{c}x_e\\y_e\\z_e\end{array}\right)
\end{equation}

The back projection of a correctly indexed point matches with the location $(i_S,j_S)_{C_2}$ of the intensity-signal on the chip of the second camera (see also Fig.~\ref{StereoMethod.eps}):

\begin{equation}
\left(\begin{array}{c}i_B\\j_B\end{array}\right)_{C_2} = \left(\begin{array}{c}i_S\\j_S\end{array}\right)_{C_2}
\end{equation}

Back projections onto camera 2 which do not match with a real signal obviously correspond to false 3D points and require further processing as follows: A first (and not a bad) approach is simply to delete the outliers. As the sensor acquires hundred thousands of points, a few missing points would be acceptable. However, as explained in the last section, $\Delta z$ shrinks if the number of projected lines increases. For a line number close to the information-theoretical maximum, only the few 3D data located in a very thin layer pass the back projection-check.

Hence, a method that \textit{corrects} the index instead of just deleting the false points is highly desired. Such a procedure will expand the usable depth range to a value which is significantly larger than $\Delta z$.

\begin{figure}[t]
\centerline{\includegraphics[width=.8\columnwidth]{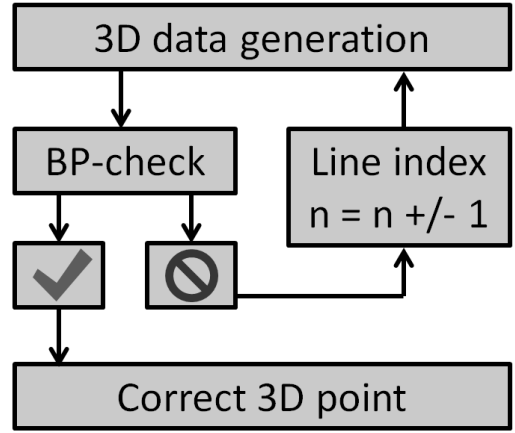}}
\caption{Scheme for the correction of false 3D points. If a 3D point fails the back projection (BP)-check, the line index is increased/decreased, and the newly evaluated 3D-point is checked again. The loop ends when the point yields correct data and passes the checkup.}
\label{CorrectionScheme.eps}
\end{figure}

Our solution consists of a trial and error increase/decrease of the line index. After evaluating a 3D-point with the new index, the back projection-check is applied again. This is done iteratively until the point passed the check. Figure~\ref{CorrectionScheme.eps} displays the method.

With the information delivered by the additional camera, a significantly larger number of lines can be used with an expanded measurement depth at the same time. The back projection-check is performed in each single camera frame for each single 3D point. Neither temporal nor lateral context information is used. Since the computational steps are simple, the method can be implemented in real time. The next section shows first experimental results. In the last section, we will discuss limits and further options.

\section{Experimental results }

The experimental proof of principle was performed in Matlab \cite{Matlab}. The 3D views and the related images of camera 2 were acquired with the current procedure, afterwards the algorithms mentioned in the last section were applied. The measurement volume of the applied multi-line triangulation sensor was $800 \times 800 \times 550  ~mm^3 $ ($\Delta x\times \Delta y \times \Delta z$) at a working distance of $1000 mm$. The sensor uses a laser light source and has a triangulation angle of $5^{\circ}$ which results in a measurement uncertainty of $1.1 mm$. Detailed information about the applied sensor can be found in \cite{Huber11,WillomitzerOSAV}.

\begin{figure}[t!]
\centerline{\includegraphics[width=\columnwidth]{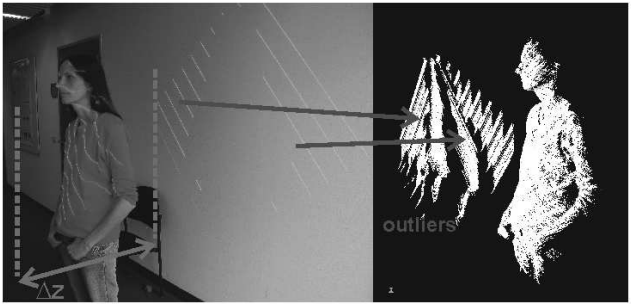}}
\caption{Test setup: Human body in front of a wall. The wall is outside the measurement range. Outliers in the 3D model are caused by false line indexing.}
\label{Exp_Evi.eps}
\end{figure}

The experiment was performed with 120 sparse 3D views of a human body. The person was placed in front of a wall, located outside the measurement depth of the sensor (see Fig.~\ref{Exp_Evi.eps}). So, the measurement volume was too small in this case. One could also say that the number of projected lines was too big to ensure unique indexing in the required (resp. actually used!) measurement  volume. The outcome is the same: with the standard data processing, the 3D data of the wall are evaluated incorrectly, due to false line indexing. In the 3D model, they can be seen as separated lines \textit{in front} of the person. Due to the severe outliers, a registration of the views takes many iterations for an acceptable result. A real time registration with this amount of outliers was not possible.

\begin{figure}[b!]
\centerline{\includegraphics[width=.8\columnwidth]{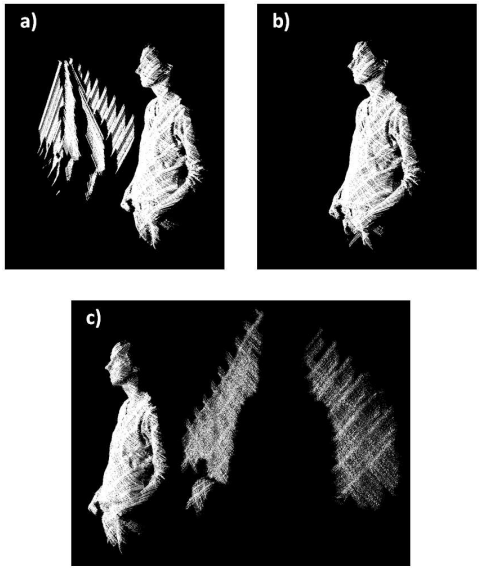}}
\caption{First implementation: sparse measurement of a human body in front of a wall which is outside the measurement range. a) Raw data: 3D points of the wall are false. b) Indentified outliers deleted, no more outliers visible.  c) Outliers corrected. The 3D data of the wall are correctly indexed and placed behind the person. The shadow of the person is visible in the 3D data of the wall.}
\label{Exp_Outcome.eps}
\end{figure}

The dataset was saved and processed with the back projection -approach. From roughly 1 Mio. acquired 3D points (Fig.~\ref{Exp_Outcome.eps}a), 250.000 were identified as false. First, they were simply deleted (Fig.~\ref{Exp_Outcome.eps}b). Second, the correction scheme of Fig.~\ref{CorrectionScheme.eps} was applied (Fig.~\ref{Exp_Outcome.eps}c). The 3D data of the wall are correctly indexed and located at the right place. After both kinds of processing, no remaining outliers are visible in the datasets.

\section{Discussion and Outlook }

\begin{figure}[b!]
\centerline{\includegraphics[width=.8\columnwidth]{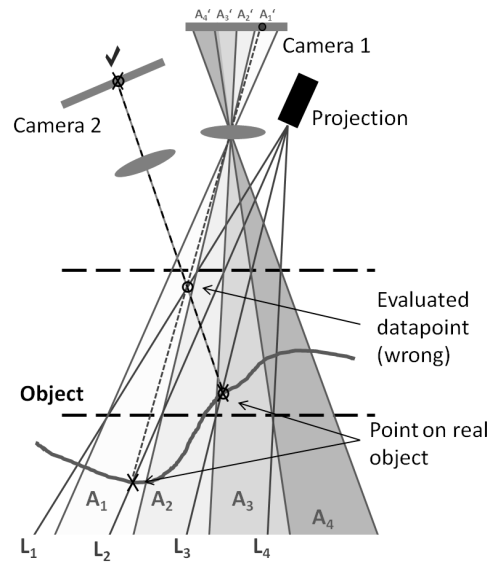}}
\caption{The back projection-approach fails, if a false indexed 3D point is back projected onto an intensity signal of a neighbored line.}
\label{Stereo_Fail.eps}
\end{figure}

The presented back projection-approach ensures the reliability of generated 3D data and at the same time increases the efficiency of multi-line triangulation systems. Without any modifications of the setup or the calibration of our sensors, the efficiency can be improved by a factor of $3\le U \le 4$, just by applying the back projection-approach. In other words, we can project 3-4 times more lines without any modification of the sensor. However, the method also has limits and fails under certain conditions. As explained above, an evaluated 3D point passes the back projection-checkup, when its corresponding pixel on the chip of the second camera is illuminated. It may happen that an incorrect 3D point is actually back projected onto the signal of a neighbored line (see Fig.~\ref{Stereo_Fail.eps}). If we consider the positions of lines and back projections on the chip of camera 2 to be random, we can calculate the likelihood $P_F$ that one false point passes the back projection-check:

\begin{equation}
 {P_F} = \frac{d_L}{D_{C}} \cdot (L-1)
\label{eq:P_F}
\end{equation}

Here, $d_L$ is the width of the line signal on the chip of the second camera in units of pixels, $D_{C}$ is the chip width perpendicular to the line direction in pixels and $L$ is the number of projected lines. For a typical multi-line triangulation setup with 15 lines and a line width of approximately $d_L = 3$ pixels, $P_F$ is roughly about 5\%. This is quite a large likelihood for a failure if we consider that only 15 lines are projected. But we will see that this statistical assumption is only an upper limit. There is much room for improvement, if the sensor fulfills certain geometrical conditions. First, reducing the factor ${d_L}/{D_{C}}$ of $P_F$ will allow to project more lines at the same fail-likelihood. Taking into account Eq.~(\ref{eq:P_F}), the only option is to reduce the cross section of the signal and the back projection. This can be done by an evaluation of the signal intensity maximum. In this case, the precise positions of signal and back projection would be compared and the fail-likelihood is reduced. However, this requires a very accurate calibration to ensure the matching of signals and corresponding back projections. Then, the upper limit of $d_L$ would be just given by the noise in the data. Hence, the development of a new calibration method which is precise and easy to execute is highly demanded. Work is in progress \cite{Schiffers14,Schroeter12}.

As already mentioned, the statistical considerations give only an upper limit for the fail-likelihood. It turns out that errors can be significantly suppressed if the position of the second camera is chosen thoughtfully. Under these conditions, the fail-likelihood is not statistical anymore.

\begin{figure}[t!]
\centerline{\includegraphics[width= 0.7\columnwidth]{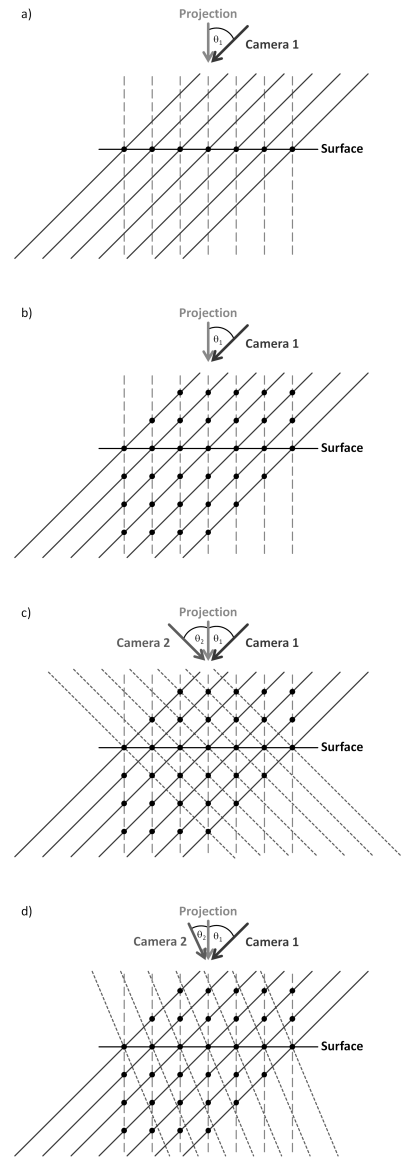}}
\caption{a) True 3D points on a surface. b) All possible candidates of 3D points due to the ambiguity problem. c) Worst position for the second camera (symetric). The gathered additive information is zero. d) One possible good position for the second camera. Only true 3D points would pass the stereo-checkup in this case.}
\label{TalbotPositions.eps}
\end{figure}

To demonstrate this, we look at the idealized case of parallel rays (telecentric setup) with a planar surface under test (according to our simulations, this is also approximately valid for slightly divergent rays and free-form surfaces). The considerations are illustrated in Fig.~\ref{TalbotPositions.eps}: The true 3D points are located at the positions where the projected lines intersect the surface under test (Fig.~\ref{TalbotPositions.eps}a). Due to potential ambiguities, all intersections of projected lines and rays of sight from camera 1 are candidates for possible 3D points (Fig.~\ref{TalbotPositions.eps}b). The key idea of the back projection-method is to add a second ray of sight from a different direction. A 3D point is then only valid if it is located on an intersection of rays from \textit{all three} directions. But where to place the second camera? The first important observation is: NOT at all symmetric to the other camera! As Fig.~\ref{TalbotPositions.eps}c shows, the ambiguities will not be resolved. A much better choice is an angle $\theta_2$ which is considerably smaller than $\theta_1$ or co prime (see Fig.~\ref{TalbotPositions.eps}d). For a very small angle $\theta_2 \approx 0$ the first remaining ambiguity lies far outside the practical measurement range. Thus, $\theta_2$ should be chosen as small as the noise allows, theoretically.

Besides the noise and the number of projected lines, there is one main practical restriction for $\theta_2$: again the quality of the calibration. Giving typical present hardware and calibration conditions, ($\theta_1 = 7^\circ$, 1Mpix camera, line width $\approx 3$~pixels),  ~$\theta_2$ has to be larger than approximately $2^\circ$, if 80 lines are projected. With this modification, uniqueness indices of $U > 10$ can be achieved. This can be also improved if, by a better calibration, the sub-pixel precise evaluation of signal position and back projections is possible.

The setup as proposed in Fig.~\ref{TalbotPositions.eps}d enables an interresting new view on the introduced back projection principle: If $\theta_2$ is chosen considerably smaller than $\theta_1$, we have in fact two sensors that use the same projection. One sensor with low precision but large unique measurement volume and one sensor with high precision and a small unique measurement volume. A unique index can be determined from the sensor with the large measurement volume. Via back projection, the identified indices are transmitted to the second sensor, which now delivers an accurate dataset. Both sensors can be combined in order to create one "perfect" sensor with high accuracy, large data density and a large measurement volume.

Beside the selection of a "good" position for the second camera, there is another possible hardware-based improvement: adding a third camera under a different angle. In this case, a 3D point will only pass the back projection-checkup if \textit{all three} cameras consider it to be correct. Detailed considerations as well as experiments for both proposed possibilities mentioned above will be published in further papers.

From a certain point, a high line density has interesting consequences: as mentioned in the third section, the initial unique depth of measurement, where 3D points are correctly evaluated, gets smaller if the number of lines increases. For a large number, this volume shrinks to a thin layer. Then, the task of generating a correct 3D point becomes exclusively a task of indexing!

A measurement principle that picks up exactly this perception is the "Tomographic Triangulation" \cite{GHTomoDGaO}. Under certain conditions, the "Tomographic Triangulation" is able to acquire pixel dense data in the time of one single camera shot, without using lateral context information and independently from ambient light. It can be interpreted as an extended back projection-approach with many (up to 15) cameras.

\section{Summary and Conclusion }

We introduced a novel method to settle the ambiguity problem in multi-line triangulation systems. The method is single-shot and does not exploit lateral context information on the camera chip. Instead, a second camera is used to resolve the ambiguities in line indexing via back projection of the evaluated 3D points. The problem was motivated and the solution approach was derived by the particular limitations of our formerly developed measurement principle Flying Triangulation. However, the resulting back projection method is not limited to a certain measurement principle. It can be applied for all multi-line triangulation systems in general. Possibly, a modification of this method is also able to resolve ambiguities in other measurement principles (e.g. fringe projection or stereo photogrammetry). 
The experiments shown prove exemplarily that the effective measurement volume of a triangulation sensor can be enlarged significantly.
As measurement depth, number of projected lines and the triangulation angle are related via the introduced "uniqueness-index", the back projection-approach can be used to increase all three quantities at the same time, creating a single-shot sensor with a large measurement volume, a high data density and a high precision: our desired 3D movie camera.


\end{document}